\documentclass[twoside,twocolumn]{article}

\usepackage{tabulary,graphicx,times,caption,fancyhdr,amsfonts,amssymb,amsbsy,latexsym,amsmath}
\usepackage[utf8]{inputenc}
\usepackage{url,multirow,morefloats,floatflt,cancel,tfrupee,textcomp,colortbl,xcolor,pifont}
\usepackage[nointegrals]{wasysym}
\usepackage{subfig}
\usepackage{svg}
\usepackage{comment}
\usepackage{hyperref}
\makeatletter

\usepackage{ifxetex}
\ifxetex\else
  \usepackage{dblfloatfix}
\fi

\@ifundefined{subparagraph}{
\def\subparagraph{\@startsection{paragraph}{5}{2\parindent}{0ex plus 0.1ex minus 0.1ex}%
{0ex}{\normalfont\small\itshape}}%
}{}

\def\URL#1#2{\@ifundefined{href}{#2}{\href{#1}{#2}}}

\def\UrlOrds{\do\*\do\-\do\~\do\'\do\"\do\-}%
\g@addto@macro{\UrlBreaks}{\UrlOrds}

\makeatother

\usepackage[paperheight=11in,paperwidth=8.3in,margin=2.5cm,headsep=.7cm,top=2.5cm]{geometry}
\usepackage[T1]{fontenc}

\renewenvironment{abstract}
	{\trivlist\item[]\leftskip0pt\par\vskip4pt\noindent
  	\textbf{\abstractname}\mbox{\null}\\}
	{\par\noindent\endtrivlist}

 \date{} \emergencystretch 8pt

\makeatletter
\def\author#1{\gdef\@author{\hskip-\tabcolsep%
	\parbox{\textwidth}{\raggedright\bfseries#1\\[1pc]}}}
\def\address[#1]#2{\g@addto@macro\@author{\\\hskip-\tabcolsep\parbox{\textwidth}{\raggedright%
	\normalsize\normalfont\textsuperscript{#1}#2}}}
\let\addresslink\textsuperscript
\def\correspondence#1{\g@addto@macro\@author{\\\hskip-\tabcolsep\parbox{\textwidth}{\raggedright%
	\vspace*{10pt}\normalsize\normalfont~\\#1~\\[12pt]}}}
\def\email#1{\g@addto@macro\@author{\\\hskip-\tabcolsep\parbox{\textwidth}{\raggedright%
	\normalsize\normalfont Emails: #1}}}

\def\title#1{\gdef\@title{\vspace*{-30pt}%
	\raggedright\textbf{\@journaltitle}~\\%
  \raggedright\bfseries\ifx\@articleType\@empty\vspace*{20pt}\else%
  \vspace*{20pt}\@articleType\vspace*{20pt}\\\fi#1}}
\let\@journaltitle\@empty \def\journaltitle#1{\gdef\@journaltitle{{\normalfont\itshape#1}}}
\let\@articleType\@empty \def\articletype#1{\gdef\@articleType{{\normalfont\itshape#1}}}

\let\@runningHead\@empty \def\RunningHead#1{\gdef\@runningHead{{\normalfont #1}}}

\usepackage{fancyhdr}
\fancypagestyle{headings}{\fancyhf{}
  \fancyhead[R]{\itshape\@runningHead}
  \fancyfoot[C]{\thepage}}
\fancypagestyle{plain}{%
	\fancyhf{}\fancyhead[R]{\LaTeX\ template for Hindawi journal articles}
  \fancyfoot[C]{\thepage}}
\makeatother

\usepackage[%
	numbers,sort&compress%
  ]{natbib}

\usepackage{float,xcolor}

\journaltitle{International Journal of Antennas and Propagation}
\articletype{Research Article} 

\begin{document}

\title{Spectral Smoothness of Ground Plane Backed Log-Periodic Dipole Antennas for Radioastronomical Applications}

\author{%
		Georgios Kyriakou\addresslink{1,2},
  	Pietro Bolli\addresslink{1} and
  	Mirko Bercigli\addresslink{3}
    }
		
\address[1]{INAF-Arcetri Astrophysical Observatory, Largo Enrico Fermi 5, 50125, Florence Italy}
\address[2]{Physics and Astronomy Department, University of Florence, Via Sansone 1, 50019, Sesto Fiorentino, Italy}
\address[3]{Ingegneria dei Sistemi, Via Enrica Calabresi 24, 56121, Pisa, Italy}

\correspondence{Correspondence should be addressed to Georgios Kyriakou: georgios.kyriakou@inaf.it}

\email{georgios.kyriakou@inaf.it, pietro.bolli@inaf.it, m.bercigli@idscorporation.com}%

\twocolumn[
\begin{@twocolumnfalse}
  \maketitle
  \begin{abstract}
  The spectral smoothness properties of the low-frequency array of the Square Kilometer Array (SKA), namely SKA-Low, are an important issue for its scientific objectives to be attainable. A large array of 256 log-periodic dipole antennas, installed on top of a 42~m circular ground plane, will work as an SKA-Low station in the frequency range 50-350~MHz. In this article, the ground plane induced effects are examined in terms of antenna beam spectral characteristics, while different antenna placements are considered. Results are produced both at isolated antenna and at array level in the band 50-100~MHz, by employing an approximate method for the speeding-up of array simulations. We attempt to distinguish the ground plane effect from that of mutual coupling among antennas, which appears to be more severe at specific frequencies, using 2 figures of merit. The Discrete Fourier Transform (DFT) components of gain pattern ratios identify the fundamental spatial components of the ripple, while the Envelope Correlation Coefficient quantifies the penalty to considering an infinite ground plane.
  \end{abstract}
\end{@twocolumnfalse}
]

\section{I.~Introduction}
A log-periodic antenna is in general one that is designed such that frequency-independent characteristics are approached. In practice though, the gain response of a log-periodic antenna, especially at the extreme of its operating band where the front-to-back ratio can deteriorate, can be enhanced by adding a ground plane that acts as a reflector. The presence of a metallic ground plane produces distinct spectral features when illuminated by an arbitrary antenna, due to the reflection from the backscattering of the antenna and diffraction from the edges.

Our main objective in this contribution is to examine the effect of a ground plane structure for a SKALA4.1 \cite{bolli2020} array. SKALA 4.1 is a dual-polarized, active log-periodic antenna that has been selected as the array element of SKA-Low. This will be the low frequency array of SKA, which in its full extent will enumerate 512 sub-arrays called stations, each composed by 256 antennas, and aims to be the largest and most sensitive radiotelescope ever built. While the principal reason for the employment of such a ground plane in SKA-Low is the antenna gain enhancement at low-frequencies, this choice also has to do with the soil underlying the antennas, which makes the observation less sensitive to the environmental parameters (moisture, mineral content etc.). 

In the first \cite{benthem2021} engineering prototype of SKA, certain concerns have been explored specific to the spectral smoothness of the antenna beam, in order to enable wide-band beamforming, as well as to fulfill the scientific prerequisite of decoupling the instrumental response from the astronomical signal of interest. On-site deployment and operation also demonstrated the calibration as well as commissioning challenges relating to large low-frequency arrays. The second prototype, Aperture Array Verification System 2.0 (AAVS2.0) \cite{macario2022}, is an array of 256 SKALA4.1 antennas with a circular ground plane 42~m in diameter, and was deployed in the Australian outback in 2019 working as a precursor of the first generation of SKA-Low stations. 

Numerical EM simulations are fundamental tools to assist in calibration routines, to characterize the sensitivity etc. \cite{bolli2022,sokolowski2021}. These simulations usually assume an infinite ground plane in order to ease the computational burden, an approximation which is in general valid for high frequencies (and thus electrically large structures). Simulations including the finite ground plane, have only been performed in some recent works \cite{cavillot2020,bolli2020_2,cavillot2022}, with the limitation of examining the antenna pattern in certain selected frequencies or just the isolated antenna. {In this work, we emphasize on the spectral properties of the ground plane induced effects examining zenith gain patterns over frequency.}

We focus here on the low-frequency (50-100~MHz) spectral response, since the diffraction phenomenon is stronger at lower frequencies \cite{collin1985}, and these are also more important for the scientific objectives of SKA-Low \cite{labate2017}. Both an isolated SKALA4.1 antenna placed at various positions with respect to its finite ground plane, as well as the Embedded Element Patterns (EEPs) of selected antennas of the full-station layout are examined, and their results are compared to the infinite ground plane solution. Two numerical solvers are used for initial verification, and the results are shown to be spectrally consistent with the far-field radiation by an arbitrarily illuminated circular plane. {In this study, the antennas and ground plane are assumed to be perfect electric conductors, while the dielectric soil volume underlying the ground plane is omitted. As examined in other works \cite{bolli2020_report,cavillot2022}, the spectral impurities due to ohmic losses are insignificant for all but the lateral antennas of a ground-plane backed array and at the lower extreme of our frequency band, while at the same time including them is computationally expensive. For these reasons, gains are identical to directivities, and the radiation efficiency is unit.} 

{\hyperref[sec:methods]{Section~II} introduces a theoretical model of the point-source induced diffraction on a circular ground plane, as well as the software used for numerical simulations of more complex antennas on a ground plane. \hyperref[sec:results]{Section~III} presents numerical simulation results, starting from a single antenna on a ground plane. A full SKA-Low station is then examined at the EEP level, as well as for its station beam. \hyperref[sec:conclusions]{Section~IV} outlines the conclusions of our analysis.}

\section{{II.~Methodology}\label{sec:methods}}

\subsection{Theoretical approach: point source illumination of a circular plane}

A circular plane of radius \( a \) illuminated by an azimuthally symmetric electric field amplitude vector \( \vec{E}_i(\rho) \) (i.e., no \( \phi \) dependence), produces a radiated electric far-field pattern \cite{collin1985}:

\begin{equation}
\hat{\theta}\cdot\vec{E}_a(k,r,\theta)=\frac{k\cos(\theta)}{r}\int_0^a\hat{\rho}\cdot\vec{E}_i(\rho)J_0(k\rho 
    \sin(\theta))\rho d\rho
    \label{eqn:hankel}
\end{equation}

Here \( k=2\pi f/c_0 \) is the wavenumber, \( f \) the frequency, \( r \) the radial distance from the reference (center of the ground plane) to the observation point while \( \vec{\rho}=\rho\hat{\rho} \) is the planar radial distance vector from the reference to the source point on the ground plane, \( \hat{\theta} \) is the zenith angle unit vector and \( J_0(\cdot) \) the zero-th order bessel function. This corresponds to a Hankel transform, but the finite upper limit means that there is a rectangular window function of size equal to the radius of the circular plane, which will cause the ground plane spatial radiation to appear in the frequency domain. If \( x_n \) is the \( n \)-th zero of \( J_0(k\rho \sin(\theta)) \), then for \( \lambda<x_n/(2\pi a) \) all 1 through \( n \) spatial tones are in the range of the integral, which means that at higher frequencies the more contributions decay into a smoother function.

\begin{figure}[h!]
    \centering
    \subfloat[]{\includegraphics[width=0.47\textwidth, keepaspectratio]{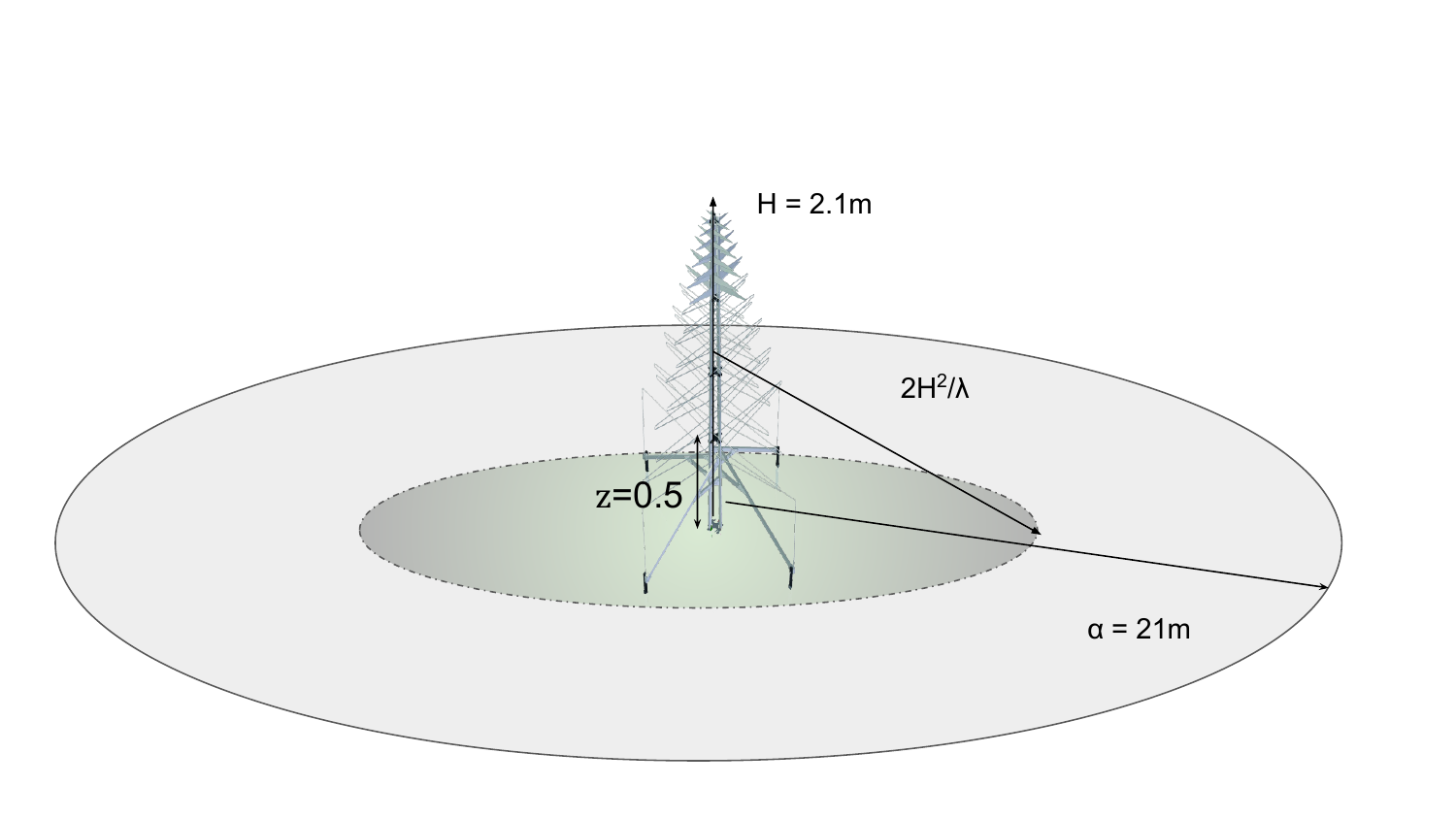}\label{fig:SKALA_groundplane_schematic}}
    \quad
    \subfloat[]{\includegraphics[width=0.475\textwidth, keepaspectratio]{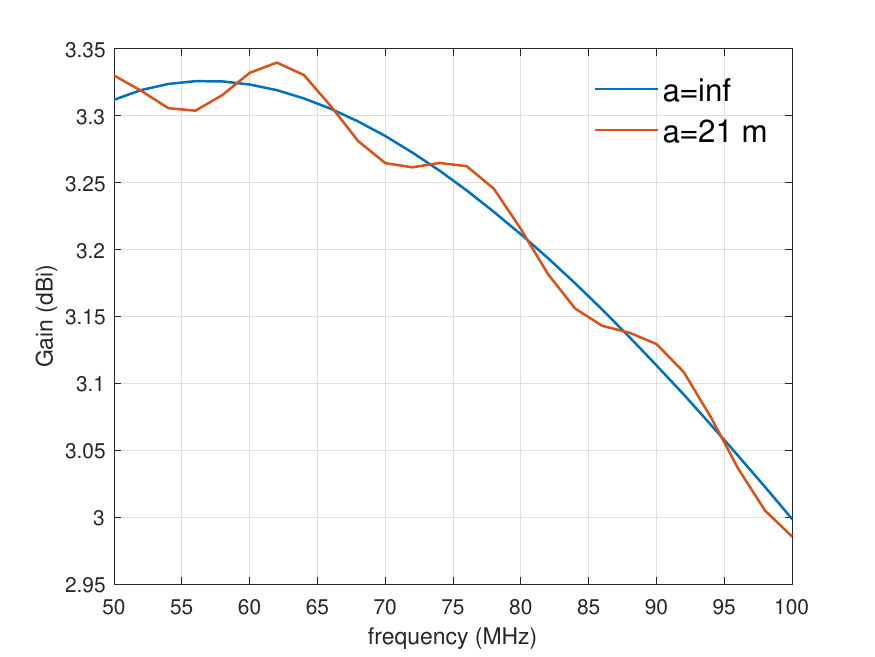}\label{fig:gain_pointsource}}
    \caption{(a): SKALA4.1 geometry, centrally placed on a 42~m diameter ground plane, which is colored differently at its intersection with the near-to-far field shpere (not at scale), (b): Gain at zenith in dBi, radiation by a circular ground plane of infinite and finite radius \( a=21\ m \) illuminated by an isotropic shperical wave source at height \( z=0.5\ m \).}
\end{figure}

We will use an isotropic spherical wave source at \( z=0.5\ m \), as a theoretic example for the illumination of the ground plane, since the diffraction effect we want to examine is in the far-field of the antenna where the illumination is roughly dependent on \( 1/\rho \). We also ignore its phase term for simplicity. Such a problem corresponds to a simplification of the configuration of \hyperref[fig:SKALA_groundplane_schematic]{Figure~1a}, whereby the phase center of the SKALA4.1 antenna is assumed to radiate as such a point source and the near-field illumination is ignored. Gain patterns at zenith of the total radiated field \( \vec{E}_{t}=\vec{E}_i+\vec{E}_a \) (taking also into account the direct isotropic radiation by the spherical source on the upper hemisphere) were calculated for this case, utilizing Eq.~(\ref{eqn:hankel}). 

The spectral gain response at zenith (in dBi) by using \( a=21\ m \), as well as an infinite ground plane, illuminated by a case of a spherical wave source at \( z=0.5\ m \) can be seen in \hyperref[fig:gain_pointsource]{Figure~1b}, for 50-100~MHz. We can see that the finite ground plane solution indeed oscillates around that of the infinite, while a negative slope is also recognized overall. This has to do with the slow reflection phenomenon of the ground plane, as outlined in \cite{bolli2020}. The gain amplitude is enhanced from 0~dBi (isotropic) to more than 3~dBi, since the ground plane is radiating strongly at zenith in this frequency range. 

An off-axis illuminated circular plane produces similar diffraction patterns with a more complicated, \( \phi\)-dependent functional form and oscillation period. Such expressions in the Fraunhofer region can be found in \cite{sheppard2004}; their implementation though exceeds the scope of this paper.

\subsection{{Numerical simulation software}}

{In this study,} we use FEKO\footnote{\url{https://altairhyperworks.com/feko/}} and Galileo\footnote{\url{https://www.idscorporation.com/pf/galileo-suite/}}, both commercial software based on a Method of Moments (MoM) solver. {FEKO uses a full-wave approach, while in Galileo an equivalent current source model has proven useful in speeding up the array simulations.}

{Particularly in the case of SKA-Low}, a large station of 256 antennas with a 42~m diameter ground plane renders the computational problem prohibitive for a full-wave solution, so we use a 3-step approximate method already known to speed up the analysis of large electromagnetic structures without significantly compromising accuracy \cite{bolli2020_2}. This method is based on the approximation of decoupling the computation of currents between antennas and ground plane, and can be summarized as follows:

\begin{enumerate}
    \item The currents on the array are computed with a full-wave solution of the MoM, using an infinite ground plane by means of its reflection
    \item These currents are then used to illuminate the finite ground plane, thus decreasing the dimensionality of the problem to the unknowns of the now meshed plane
    \item The radiated electric field components of both structures are superimposed
\end{enumerate}

\section{{III.~Results \label{sec:results}}}

\subsection{Numerical simulations of SKALA4.1}

For an antenna such as SKALA4.1, it is expected that when placed at a central position and illuminating the ground plane, a ripple of the same periodicity as the theoretic example will appear. For off-center antenna placement on the ground plane, the illumination is more complex, the geometry is not azimuthally symmetric and we cannot intuitively expect the presence of the same ripple. We must also note that SKALA4.1 has an electrical connection to the ground plane and does not always illuminate in the far-field (see schematic of \hyperref[fig:SKALA_groundplane_schematic]{Figure~1a}), which makes the results of a full-wave solver different than if we attempted to use the SKALA4.1 far-field pattern as an illumination in Eq.~\ref{eqn:hankel}.

\begin{figure}[h!]
    \centering
    \subfloat[]{\includegraphics[width=0.47\textwidth, keepaspectratio]{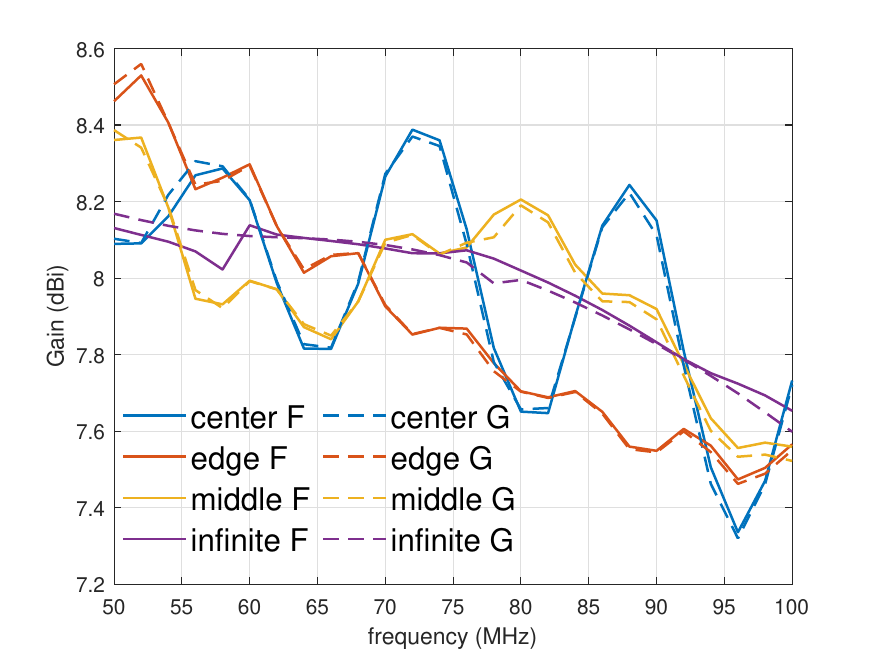}\label{fig:Gain_isolatedants}}
    \quad
    \subfloat[]{\includegraphics[width=0.475\textwidth, keepaspectratio]{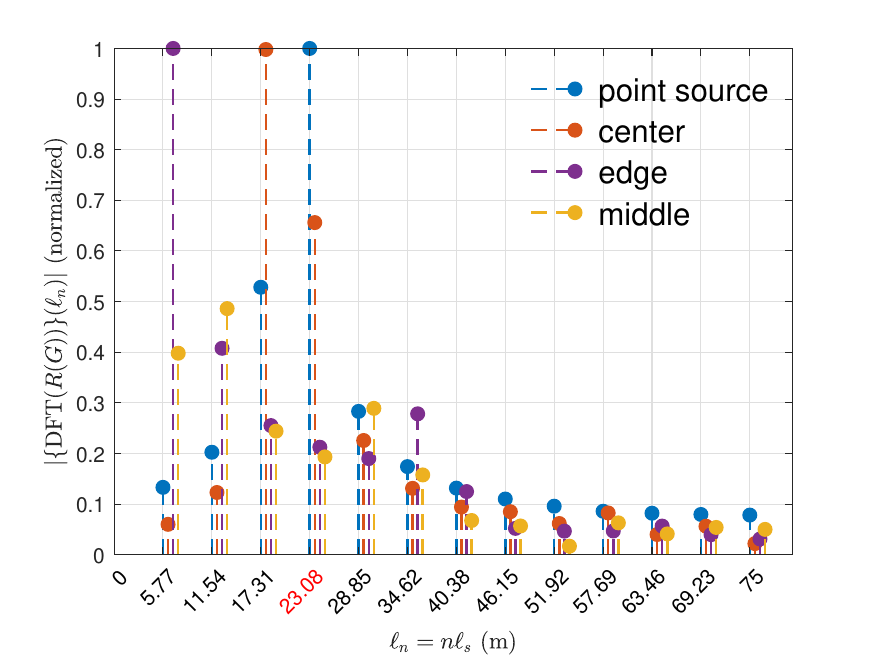}\label{fig:DFT_isolatedants}}
    \caption{(a): Simulated gain at zenith in dBi, of a SKALA4.1 antenna placed on 3 positions of both infinite and finite ground plane cases, using FEKO ("F" in label) and Galileo ("G" in label), (b): DFT amplitudes of the gain ratio \( R(G) \) between infinite and finite ground plane solution patterns, for all 3 isolated antennas of \hyperref[fig:aavs2_topdown]{Figure~3} and the theoretic point source illumination. The sampled lengths are \( \ell_n=n\ell_s \), and we also normalize the vertical axis to its maximum.}
    \label{fig:DFT}
\end{figure}

As our first step in this numeric analysis, we simulate in \hyperref[fig:Gain_isolatedants]{Figure~2a}, the pattern of a SKALA4.1 placed on a finite \( a=21\ m \) and infinite radius ground plane, and excite its Y-polarization. Moreover, to test the position dependence, we choose three positions, termed "center", "middle" and "edge", corresponding to the positions of antennas \#120, \#132, \#130, respectively, of the next section (see \hyperref[fig:aavs2_topdown]{Figure~3}, but isolated on the ground plane for this section), and focus on 50-100~MHz with a 2~MHz step. 

In \hyperref[fig:Gain_isolatedants]{Figure~2a}, we first notice the gain at zenith results for the centrally placed antenna of either solver (blue curves). We can recognize a qualitatively similar ripple as in the theoretic spherical wave source case, as far as periodicity is concerned, although it is seen here to have a larger amplitude (since SKALA4.1 is more directional by design). There is again a negative slope owing to the reflection phenomenon by the ground plane of the antenna backscattering, which has initially been observed in the infinite ground plane simulations, as analysed in \cite{bolli2020}. 

We then examine the gain response at zenith for all the antennas chosen, including comparisons between the two solvers used. In \hyperref[fig:Gain_isolatedants]{Figure~2a}, it can be observed that all finite ground plane patterns exhibit more complex periodic deviations from the infinite ground plane solution, indicating that the contribution of the ground plane across frequency is not limited at the zero-th order Hankel wavenumber. The two solvers give similar results. It is also evident that the periodicity away from the center is not the same, and not always centered around the infinite ground plane curve (e.g., the edge antenna has a systematic offset).

To look for the existence of periodic characteristics, we will use a Discrete Fourier Transform (DFT) on the data\footnote{The most appropriate would be a Discrete Hankel Transform, but since the data is band-limited, the complexity of this method to interpret such data is higher, so it will be the objective of future studies. Our implementation here was based on the relevant Matlab function utility \url{https://ch.mathworks.com/help/signal/ug/discrete-fourier-transform.html}}. Since we are using a sampling rate of \( f_s=2\ \text{MHz} \) and \( N=26 \) frequency points, the spatial-domain sampling will be \( \ell_s=\frac{c_0}{Nf_s}=5.77\ \text{m} \). By applying the DFT on real-number data as are the gain ratios between infinite and finite ground plane solution data (\( R(G) \) on the \hyperref[fig:DFT_isolatedants]{Figure~2b} y-axis) we get a symmetric discrete spectrum, so we keep only the one-sided result, as in \hyperref[fig:DFT_isolatedants]{Figure~2b}. We then normalize these amplitudes with respect to maximum so that the point source illumination results is comparable to the SKALA4.1 simulations. 

We can now assess how close the amplitude peaks of each case approach \( L=23.07\ \text{m} \), which is the fundamental length that the point source solution peaks at (as expected, close to our radial length \( a=21\ \text{m} \) highlighted in red), or what multiple of this fundamental length they correspond to, by examining the discrete spectrums of \hyperref[fig:DFT_isolatedants]{Figure~2b}. It can be seen that for the central antenna the highest peak is present at \( 0.75L \), while the one which deviates the most is the edge antenna, whose fundamental peak lengths seem to be \( 0.25L \) and \( 0.5L \) (descending order), indicating that the distance from the edge is interpreted as the effective radius "seen" by the diffraction integral. Finally, the middle antenna has 2 peaks at \( 0.5L \) and \( 0.25L \) (descending order), leading to the same conclusion for its diffraction properties.

\subsection{Emdedded Element Patterns {of a SKA-Low station}}

\begin{figure}[h!]
    \centering
    \includegraphics[width=0.47\textwidth, keepaspectratio]{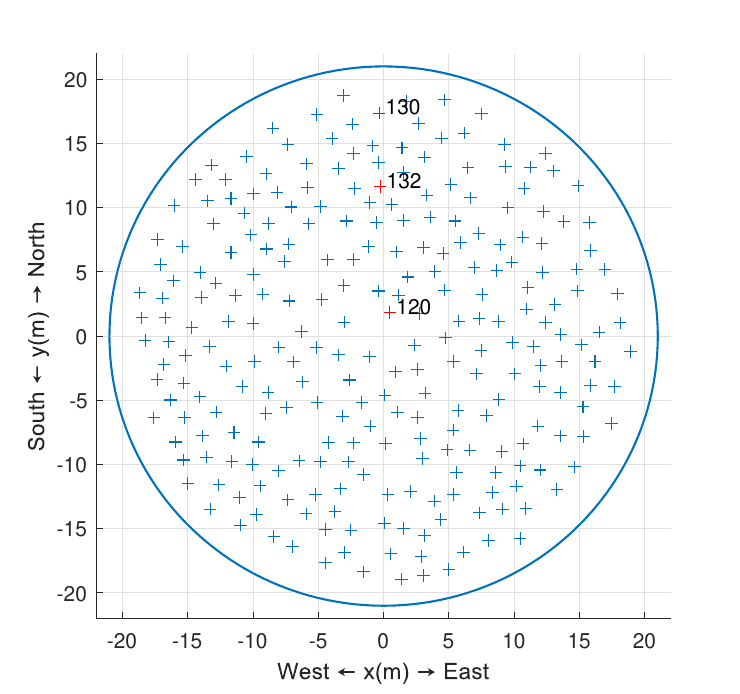}
    \caption{AAVS2.0 two dimensional layout and ground plane perimeter, highlighted by crosses are all the antennas, while the examined ones are colored red and their identifying number is given.}
    \label{fig:aavs2_topdown}
\end{figure}

As mentioned above, a SKA-Low station consists of 256 SKALA4.1 antennas, arranged in a quasi-random configuration as in \hyperref[fig:aavs2_topdown]{Figure~3}, which shows the AAVS2.0 layout. The pattern of each antenna when excited, while all other antennas are passive, is an Embedded Element Pattern (EEP) and differs from others due to the antenna mutual coupling being irregular. We simulated 3 such patterns in the same frequency range as before, exciting them with unit amplitude and a \( 50\ \Omega \) source impedance while also terminating the passive antennas with \( 50\ \Omega \).

\begin{figure}[h!]
    \centering
    \subfloat[]{\includegraphics[width=0.475\textwidth, keepaspectratio]{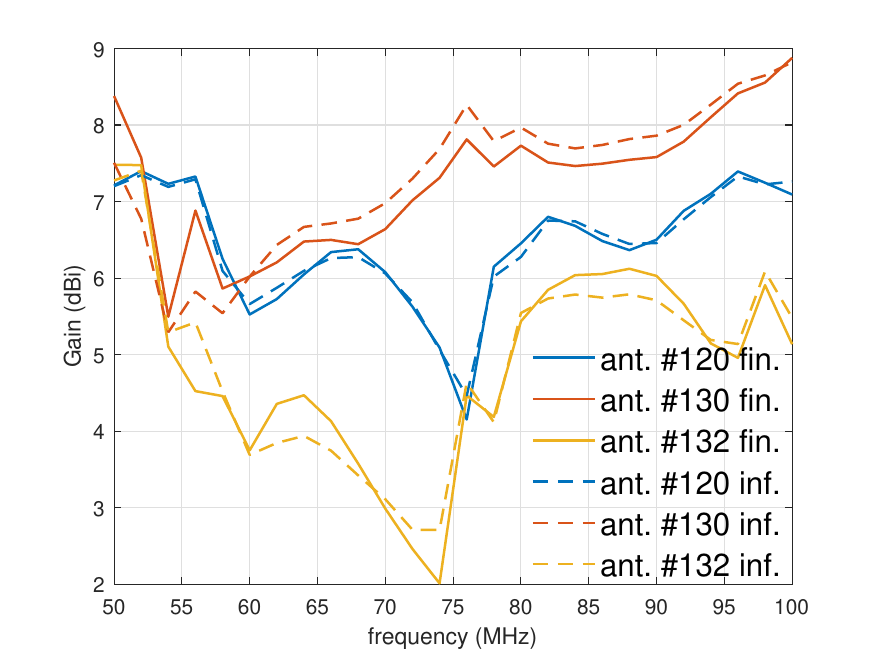}\label{fig:Gain_eeps_groundplane}}
    \quad
    \subfloat[]{\includegraphics[width=0.475\textwidth, keepaspectratio]{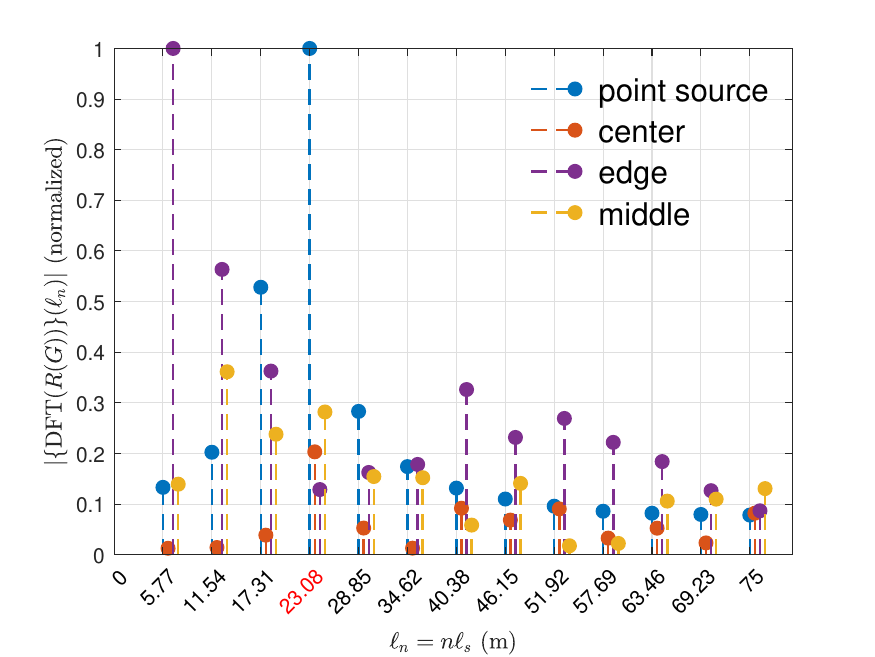}\label{fig:DFT_eeps}}
    \caption{(a): Simulated gain at zenith of EEPs in AAVS2.0 using Galileo, for all 3 antennas of \hyperref[fig:aavs2_topdown]{Figure~3}. Included here are also the same quantities for the infinite ground plane solution, (b): Same as \hyperref[fig:DFT_isolatedants]{Figure~2b}, but for the embedded antenna patterns and the same theoretic point source data.}
    \label{fig:embedded_EEC}
\end{figure}

\hyperref[fig:Gain_eeps_groundplane]{Figure~4a} presents the selected EEP gains at zenith for the full-station with a finite ground plane in comparison to the corresponding ones computed with an infinite ground plane. We can see in this plot that the mutual coupling disturbs the patterns, with respect to the single antenna patterns seen in \hyperref[fig:Gain_isolatedants]{Figure~2a}, across the whole frequency range, with a number of mutual coupling related gain drops of 2-3~dB in magnitude. Usually, the EEP distortion is averaged out when forming the beam from all of them, however there are frequency regions where glitches of not random nature occur and therefore they are still present in the station beam as investigated in \cite{bolli2022_2}, {where a detailed analysis with an infinite ground plane showed the limitations of the SKA-Low station in the dense regime}. Our focus here is rather on the difference between finite and infinite ground plane cases: these are less than 1~dB and this difference is fluctuating between positive and negative values, {while the general EEP trend determined by mutual coupling remains qualitatively the same}. To acquire an estimate of any periodicity on the ratio between finite and infinite ground plane cases, we present the normalized DFT amplitudes as has been done in the isolated case, in \hyperref[fig:DFT_eeps]{Figure~4b}. The central antenna, \#120 in AAVS2.0 now peaks at the fundamental length \( L \), even with some slight peaks elsewhere. Antennas \#130 and \#132, as with their isolated "middle" and "edge" counterparts, have a more complex shape: notably, we identify that both of them have a much more spread-out spectrum even if they mostly retain their peak components inherited from the isolated antennas.
 
\begin{figure}[h!]
    \centering
    \includegraphics[width=0.475\textwidth, keepaspectratio]{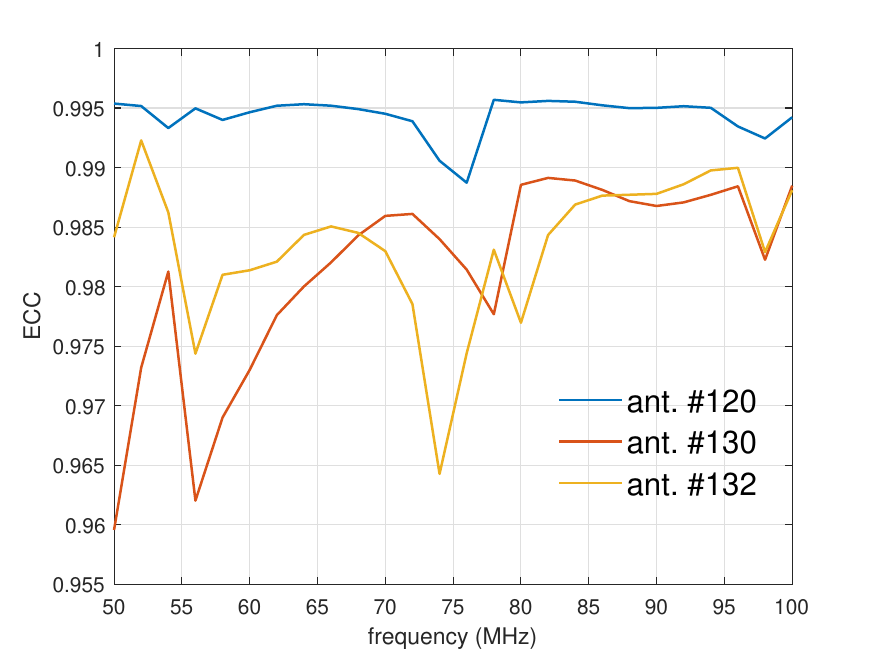}
    \caption{Envelope Correlation Coefficient vs. frequency for the same three AAVS2.0 embedded antennas.}
    \label{fig:ECC}
\end{figure}

Next we also present the Envelope Correlation Coefficient (ECC), calculated between the infinite (\( \vec{E}_1(\theta,\phi) \)) and finite (\( \vec{E}_2(\theta,\phi) \)) ground plane solutions. This coefficient is defined as:
\begin{equation}
    \begin{array}{c}
     {\rm ECC}(\vec{E}_1,\vec{E}_2)=\\ \\
     \frac{\left|\int_0^{2\pi}\int_0^{\pi}\vec{E}_1(\theta,\phi)\cdot\vec{E}_2^*(\theta,\phi)d\theta d\phi\right|^2}{\int_0^{2\pi}\int_0^{\pi}|\vec{E}_1(\theta,\phi)|^2d\theta d\phi\int_0^{2\pi}\int_0^{\pi}|\vec{E}_2(\theta,\phi)|^2d\theta d\phi}
     \end{array}
\end{equation}

and makes use of the full complex valued electric field patterns \( \vec{E}_1,\ \vec{E}_2 \) over the 3D sphere to evaluate a correlation-like metric. For the selected antennas of the AAVS2.0 station, we calculate this coefficient using both electric field \( E_{\theta} \) and \( E_{\phi} \) quantities and present the results in \hyperref[fig:ECC]{Figure~5}. Antenna \#120 seems to best follow its infinite ground plane performance. In contrast, the middle and edge antennas have the most disturbed coefficients. It is noteworthy though that all three EEC drops are centered around the mutual coupling glitches identified at 55 MHz, 77 MHz in \cite{bolli2022_2}, and 98 MHz, showing that this effect can be volatile depending on the ground plane size. Due to the complexity of both the ground plane ripple and the mutual coupling phenomena, they cannot be isolated and quantified separately in terms of ECC, but these values are always over 0.96 so the amplitude of each EEP beam volatility with respect to the ground plane being finite or infinite should be limited. 

\subsection{Station beam {of a SKA-Low station}}
In this section, we will examine the station beam for some of the frequencies in the 50-100~MHz range. We will use the EEPs to construct the total weighted electric field response. We have the station beam electric field expressed as:

\begin{equation}
    \begin{pmatrix} E_{st,\theta}^{Z_g} & E_{st,\phi}^{Z_g}\end{pmatrix}=\bar{w}_g^H\cdot\begin{pmatrix} \bar{E}_{\theta}^{Z_g} & \bar{E}_{\phi}^{Z_g}\end{pmatrix}
    \label{eq:weighting}
\end{equation}

where the overbar quantities are \( 256\times 1 \) column vectors, the superscript is used to indicate EEPs calculated with the \( Z_g=50\ \Omega \) termination and \( \bar{w}_g \) are the voltage weights. For zenith pointing, these are all unit. The radiated power \( P_{rad}^{Z_g} \) is calculated by integrating the far-field intensity over the 3D sphere as:

\begin{equation}
    P_{rad}^{Z_g}=\frac{1}{2\eta_0}\int_0^{2\pi}\int_0^{\pi}(|E_{st,\theta}^{Z_g}|^2+|E_{st,\phi}^{Z_g}|^2)r^2d\theta d\phi
    \label{eq:Prad}
\end{equation}

\begin{figure*}
    \centering
    \subfloat[55 MHz]{\includegraphics[width=\textwidth, keepaspectratio]{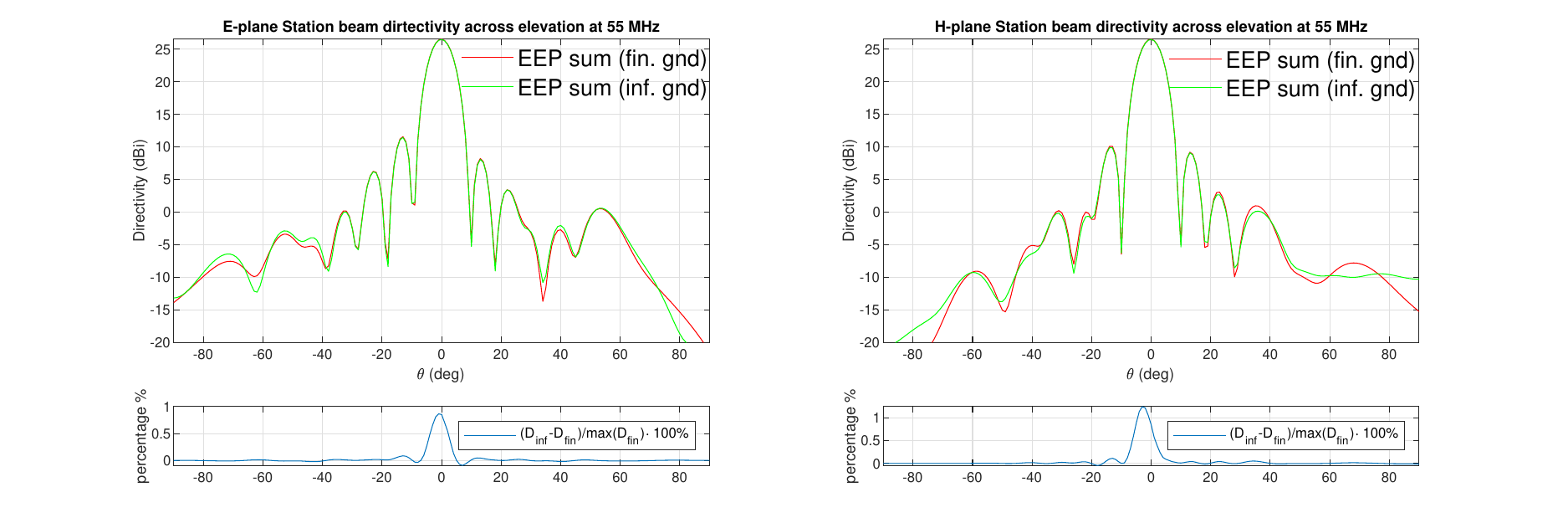}}
    \newline
    \subfloat[65 MHz]{\includegraphics[width=\textwidth, keepaspectratio]{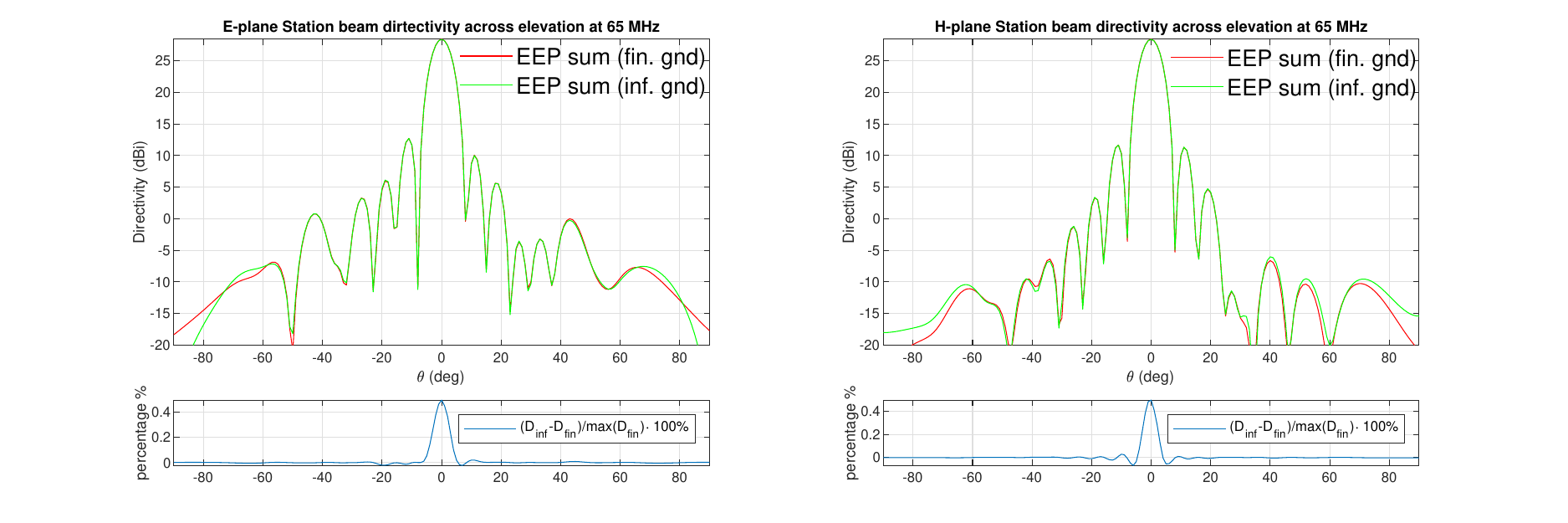}}
    \newline
    \subfloat[77 MHz]{\includegraphics[width=\textwidth, keepaspectratio]{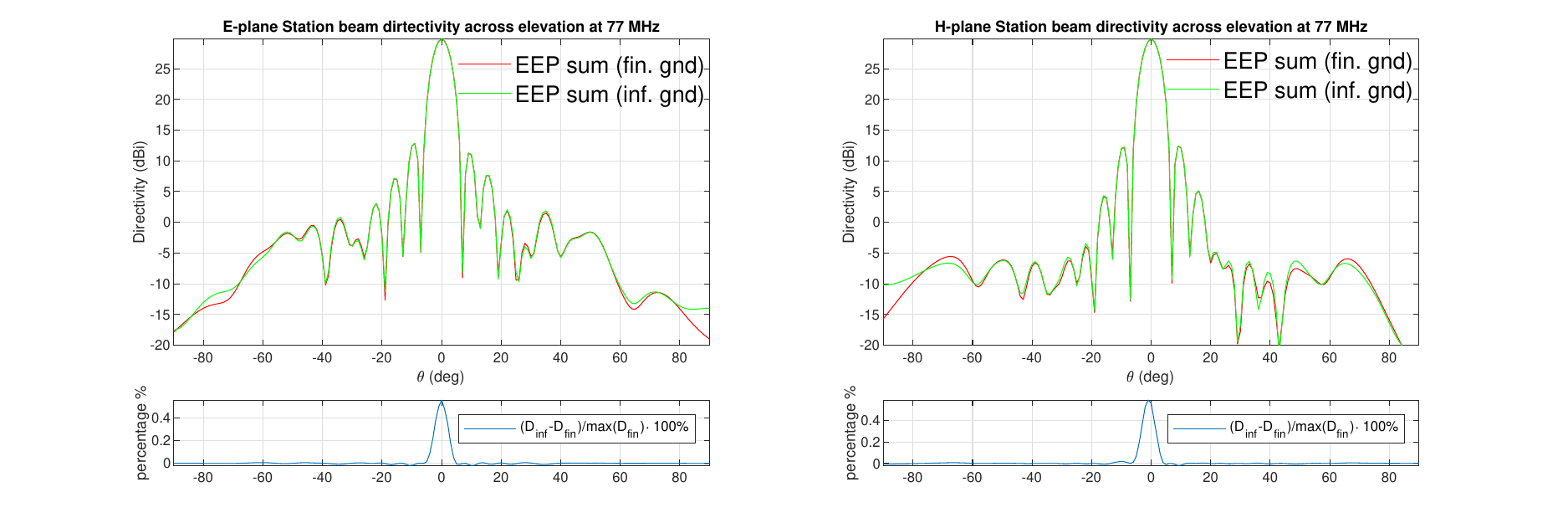}}
    \caption{Station beam directivity of an AAVS2.0 layout of SKALA4.1 antennas, at 55~MHz, 65~MHz and 77~MHz (top to bottom), for E-plane (left) and H-plane (right). The full-wave EEP summed solutions of the finite and infinite ground plane cases are presented. The bottom panels show the percentage difference, normalized to the beam maximum, of the finite and infinite ground plane cases.}
    \label{fig:station_beam}
\end{figure*}

In \hyperref[fig:station_beam]{Figure~6}, we present the station beam directivities for zenith pointing, for the E- and H- plane of an AAVS2.0 layout station using the above approach and utilizing the infinite ground and finite ground solutions at 55, 65 and 77~MHz. As far as the maximum value is concerned, at 55~MHz a deviation of 0.7~dB of the EEP solutions with respect to the array factor method is noticed. This means that the station beam is not significantly compromised by the mutual coupling at this glitch frequency, whereas at 77~MHz the deviation is even less significant. The finite ground plane station beam gains are in excellent agreement with those of the infinite ground plane solution. The lower panels present the difference between finite and infinite ground plane solutions normalized to the maximum of the infinite case, which fall off from a peak value of about 1\% at zenith.

A last useful exercise is to quantify, in terms of beamforming efficiency what the maximum deviation induced by the ground plane finiteness could be by examining the maximum-directivity beamformer. This choice is prompted by the SKALA4.1 pattern non-uniformity, as well as the mutual coupling phenomena being strong, both of which would lead to pointing errors by beamforming only with phase terms \cite{bolli2021,davidson2022}. By using Eq.~(\ref{eq:Prad}) and substituting Eq.~(\ref{eq:weighting}) (dropping the \( Z_g \) appex on the E-fields), one can arrive at the expression \cite{warnick2018}:
\begin{gather}
    P_{rad}^{Z_g}=\bar{w}_g^H\mathbf{A}\bar{w}_g \nonumber\\
    =\bar{w}_g^H\left(\int_0^{2\pi}\int_0^{\pi}\frac{1}{2\eta_0}(\bar{E}_{\theta}^{H}\bar{E}_{\theta}+\bar{E}_{\phi}^{H}\bar{E}_{\phi})r^2d\theta d\phi\right)\bar{w}_g
    \label{eq:Prad_dev}
\end{gather}

The matrix \( \mathbf{A} \) in the above equation is called the pattern overlap matrix, and it is used to calculate the maximum beamformer weights. We introduce the pedex \textit{fin/inf} for the finite or infinite ground plane, respectively. Then, by the knowledge of \( \mathbf{A}_{inf} \) one can beamform with maximum directivity towards \( (\theta_0,\phi_0) \) using the weights:

\begin{equation}
    \bar{w}_{g0,inf}(\theta_0,\phi_0)=\mathbf{A}_{inf}^{-1}\bar{E}_p(\theta_0,\phi_0)
    \label{eq:w_opt}
\end{equation}

where \( {p} \) denotes the desired polarization component. We use the Ludwig-III definition; others can be followed as well \cite{aboserwal2018}. Note that in this sense, only the partial, co-polarized directivity is maximized (the full-polarized directivity can be maximized by another, rank-2 generalized eigenvalue problem, but this is rarely a real case). Finally, we can extract the beamformer efficiency as\footnote{We take the ratio of intensities, since the normalization by the radiated power, which is equal for both cases due to the approximate method used, cancels out.}:

\begin{equation}
    \eta_{BF}(\theta_0,\phi_0)=\frac{\bar{w}_{g0,inf}^H\mathbf{B}_{fin}\bar{w}_{g0,inf}}{\bar{w}_{g0,inf}^H\mathbf{B}_{inf}\bar{w}_{g0,inf}}
    \label{eq:eta_BF}
\end{equation}

where \( \mathbf{B}=\bar{E}_{\theta}^{H}\bar{E}_{\theta}+\bar{E}_{\phi}^{H}\bar{E}_{\phi} \) is the array response matrix \cite{warnick2018}. From an analogous expression, \( \eta_{BF} \) is derived in \cite{wijnholds2020} for noisy beam weights \( \bar{w}_{g0,inf} \); in our case, the term efficiency refers to uncertainties due to the beam modelling when one uses infinite ground plane patterns instead of finite, while these weights are assumed to be perfectly known. 

\begin{figure}
    \centering
    \subfloat[55 MHz]{\includegraphics[width=0.47\textwidth, keepaspectratio]{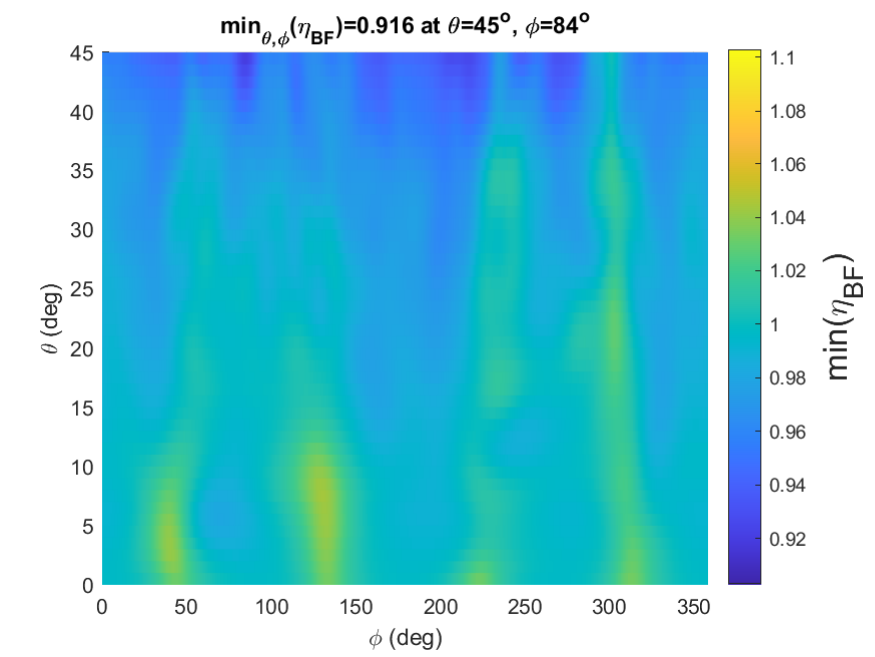}}
    \newline
    \subfloat[65 MHz]{\includegraphics[width=0.47\textwidth, keepaspectratio]{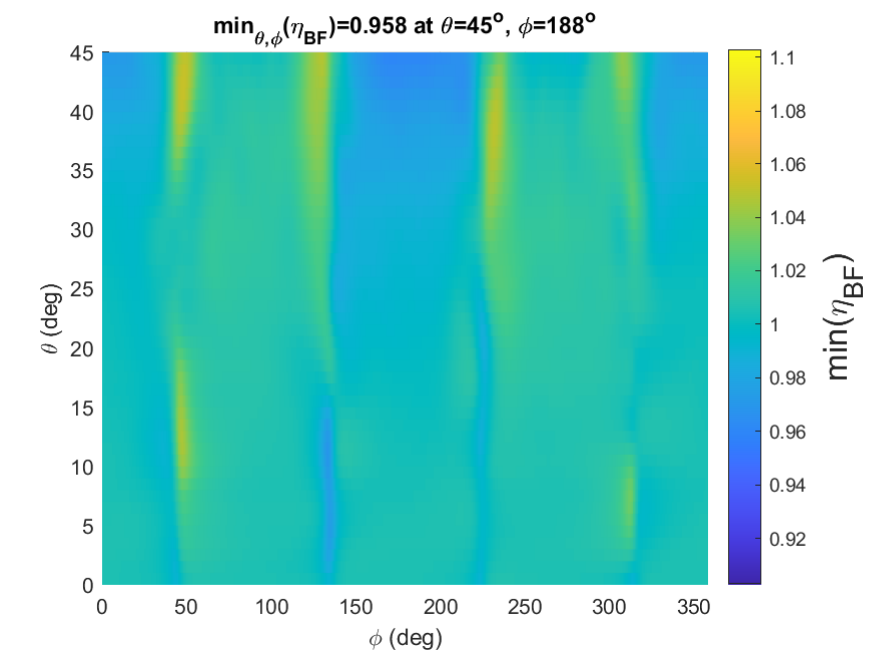}}
    \newline
    \subfloat[77 MHz]{\includegraphics[width=0.47\textwidth, keepaspectratio]{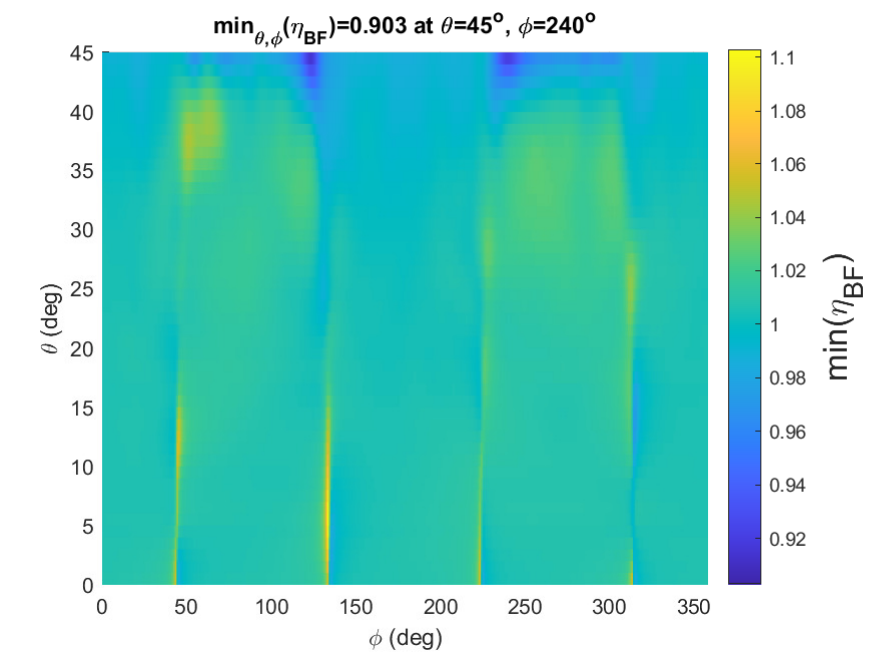}}
    \caption{Beamforming efficiency \( \eta_{BF}(\theta_0,\phi_0) \) as calculated by Eqs.~(\ref{eq:w_opt}),~(\ref{eq:eta_BF}) for all \( 0^\circ\leq \theta_0 \leq 45^\circ \), \( 0\leq\phi_0<360^\circ \), at three frequencies (55, 65 and 77~MHz). The colorbars are identical, while the titles denote the minimum and its position.}
    \label{fig:eta_BF}
\end{figure}

We solve this problem for every direction within a field of view \( \Omega=\{0^\circ\leq\theta_0\leq 45^\circ,\ 0^\circ\leq \phi_0<360^\circ \} \), and present the colormaps of \( \eta_{BF}(\theta_0,\phi_0) \) in \hyperref[fig:eta_BF]{Figure~7} for the 3 frequencies examined before; the colorbars are identical and capture the min/max over all 3 frequencies. We see that the highest values exceed 1 by about 10\%, which leads to an underestimation of the array's beamforming performance by using the infinite ground plane patterns, while the lowest ones are dwindled by about 10\% leading to a corresponding overestimation.

It is also evident from \hyperref[fig:eta_BF]{Figure~7} that, while at 55~MHz most of values are \( \eta_{BF}<1 \), the opposite trend is seen at the other frequencies, and especially at 77~MHz. This means that the ground plane information does not completely wear out with the randomized layout; the complexity of its illumination as well as mutual coupling prevent us though from performing more careful statistics on the station beam ripple. Last, we can see that the principal planes at \( \phi=45^\circ,\ 135^\circ,\ 225^\circ,\ 315^\circ \) are the central regions where most deviations are present. This effect has to do with the importance of these planes for the Ludwig-III cross-polarized pattern \cite{aboserwal2018}: for highly azimuthally uniform patterns, a cross-polarization maximum/minimum is present there, which means that beamforming with an aim to maximimze the co-polarized pattern is impacted more/less by the extra term of cross-polarization in \( \mathbf{B} \). Essentially, when \( \eta_{BF}>1 \), we use the max-directivity beamformer to our advantage by utilizing the co-polarized diffraction of the ground plane to maximize the intensity of the nominator in Eq.~(\ref{eq:eta_BF}).

\section{IV.~Conclusions \label{sec:conclusions}}

A 256-antenna SKA-Low station presents certain spectral characteristics in the 50-100~MHz band owing to the 42~m circular ground plane placed under the antennas. We identified and characterized this effect for isolated as well as embedded antennas into a station of the AAVS2.0 layout, as being related to the principal Hankel transform spatial components of the radiating circular plane or certain multiples of it. Both solvers used, FEKO and Galileo, are shown to agree and a numerical approximation method by Galileo speeds up the simulation time for the array. We also report the intertwining of this effect with the mutual coupling spectral glitches already known to appear in certain frequencies using a DFT to identify the fundamental length for isolated and embedded antenna gain pattern ratios, and the ECC to quantify the level of the combined effect. For the EEPs, the DFT components peak at different values of the sampled space of lengths, which are generally lower the closer an antenna approaches the edge of the ground plane, while their ECC penalty when calculated with respect to the infinite ground plane EEPs is always less than 4\% in the examined frequency range. For the station beam patterns, the normalized difference between infinite and finite ground plane solution is limited to 1\% for uniform beamforming, while it can suffer up to 10\% efficiency loss due to ground plane modelling errors in the worst-case maximum-directivity beamforming scenarios.

\section{Data Availability}

The antenna pattern data as well as visualization scipts used for this study can be made available under request to the corresponding author. 

\section{Conflicts of Interest}

The authors declare that they have no conflicts of interest.

\section{Acknowledgments}
The authors would like to acknowledge Ravi Subrahmanyan for having read the paper and provided useful comments on the results.

\end{document}